\documentclass[amsmath,amssymb,twocolumn,prl]{revtex4-1}

\usepackage{graphicx}               
\usepackage{bm}                     
\usepackage{hyperref}               
\usepackage{color}
\usepackage[bold]{hhtensor} 
\usepackage{multirow}

\begin{document}

\title{Velocity Slip on Curved Surfaces}

\author{Weikang Chen}
\email{wchen@ccny.cuny.edu}

\author{Rui Zhang}
\email{ruizhang@ccny.cuny.edu}

\author{Joel Koplik}
\email{koplik@sci.ccny.cuny.edu}

\affiliation{
 Benjamin Levich Institute and Departments of Physics\\ 
 City College of the City University of New York, New York, NY 10031
}

\date{\today}

\begin{abstract}
{
The Navier boundary condition for velocity slip on flat surfaces, when expressed in tensor 
form, is readily extended to surfaces of any shape.  We test this assertion using molecular 
dynamics simulations of flow in channels with flat and curved walls and for rotating 
cylinders and spheres, all for a wide range of solid-liquid interaction strengths.  We find 
that the slip length as conventionally measured at a flat wall in Couette flow is the same 
as that for all other cases with curved and rotating boundaries, provided the atomic 
interactions are the same and boundary shape is properly taken into account.  These results 
support the idea that the slip length is a material property, transferable between different 
flow configurations.
}
\end{abstract}


\maketitle

The explosive growth in the development and application of 
microfluidic devices requires accurate modeling of fluid flow in irregular and convoluted 
regions with curved bounding surfaces.  At the same time, the traditional no-slip boundary 
condition (BC) for the velocity of a liquid at a solid surface has come into question 
\cite{Stone} and 
attention has focused on alternatives, and in particular the velocity slip boundary 
condition first proposed by Navier \cite{Navier} in 1823.  In its usual form for flow past 
a flat solid surface, one introduces a slip length $\xi$ as the distance from 
the surface where the linearly-extrapolated fluid velocity field coincides with the surface 
velocity.  Explicitly, the discontinuity $\Delta V$ between the fluid and solid
tangential velocities at the surface is assumed to be proportional to the local strain rate:
\begin{equation}
	\xi\left({\partial u_x\over\partial y}\right)_S=\Delta V 
\label{eq:lbc}
\end{equation}
where $x,y$ are Cartesian coordinates parallel 
and normal to the surface $S$, respectively.  This Navier slip BC has been been widely
used in gas dynamics \cite{Zhang} since the work of Maxwell \cite{Maxwell}, and in the last decade
or two slip has been observed for liquid flows both in experiments (see the reviews in
\cite{Stone,Craig1,Granick2,Leger,Bouzigues}) and 
molecular dynamics computer simulations.  The latter have
indicated that the slip length in liquids depends critically on three factors --
wettability, roughness and strain rate. A fluid is more likely to slip in the presence
of a weak liquid-solid interaction (indicated by a high contact angle) 
\cite{Koplik,Barrat1,Barrat2,Cieplak,Huang,Rui}, surface roughness at the atomic scale 
influences the degree of slip in a complex way \cite{Sun,Craig2,Galea} 
and the slip length tends to grow and perhaps diverge at high strain rate flows 
\cite{Thompson,Priezjev,Martini}. 

Going beyond flat surfaces, in 1990 Einzel, Panzer, and Liu \cite{Einzel} introduced
a curvature correction to the Navier slip length, which was pursued in some gas 
dynamics studies \cite{Tibbs,Aoki,Wang,Dongari}, and later expressed in a general tensor
form by Barber {\em et al}. \cite{Barber}.  If one regards velocity slip as the
linear response of the fluid to the shear stress exerted at the fluid-solid interface,
then a coordinate invariant generalization of Eq.~\ref{eq:lbc} is
\begin{equation}
\label{eq:gbc}
   \frac{\xi}{\mu}\;\vec{\tau} : \vec{\hat{n}}\,\vec{\hat{t}} =\Delta V
\end{equation} 
where $\vec{\tau}$ is the shear stress tensor, $\mu$ is the fluid viscosity and $\vec{\hat{n}}$ 
and $\vec{\hat{t}}$ are normal and tangent unit vectors at the surface, respectively. 
The generalization assumes that the solid is impenetrable and the normal fluid velocity
vanishes at the surface, and the two boundary conditions agree when the surface is flat.  More generally,
as we shall see, if we choose a coordinate system ``aligned'' with the surface the curvature
corrections emerge naturally.

In this paper, we will use molecular dynamics (MD) simulations of simple liquids to test 
the above form of the Navier BC for curved surfaces. The slip length can be extracted directly
from the velocity field obtained in simulations of flow past various solid boundary shapes,
including planes, cylinders and spheres. Alternatively, we can compare the torque
on a rotating solid to the solutions of the Navier-Stokes equation with a Navier BC imposed and
 infer the slip length.  We address the influence of wettability by varying
the strength of the interaction between liquid and solid atoms, and 
the effects of roughness are
avoided by using model solids with atomically-smooth surfaces with a fixed lattice structure, 
and likewise we do not explore the variation of slip length with other solid properties such 
as the atomic mass or the stiffness of the binding potential \cite{Asproulis}. 
The key feature of the
Navier condition which is tested here is whether the slip length is a genuine material 
parameter, dependent on the nature of the solid and liquid involved but otherwise a constant
transferable between different flow configurations.  

We assume that the fluid is Newtonian and incompressible, so that the stress is proportional to the 
deviatoric stress tensor, $\vec{\tau} = 2\mu \matr{E}$.  In Cartesian coordinates, the Navier slip law
Eq.~\ref{eq:gbc} reduces to 
\begin{equation}
\label{eq:cbc}
\xi \left( \frac{\partial u_i}{\partial r_j}+\frac{\partial u_j}{\partial r_i} \right)_S t_i\,n_j = \Delta V
\end{equation}
and for a plane wall, the Cartesian coordinates can be aligned with the wall, and Eq.~\ref{eq:lbc} is
recovered. 
A trivial generalization to a curved wall would replace $\partial u_x/\partial y$ by $\partial u_t/\partial 
r_n$, but Einzel {\em et al}. Ref.~\cite{Einzel} pointed out the correct procedure is to begin with Eq.\ref{eq:cbc} 
and work out the derivatives carefully, leading to an extra term related to curvature. A simpler and more
systematic procedure is to directly evaluate the general form of the boundary condition Eq.\ref{eq:gbc} 
in a curvilinear coordinate system aligned with the boundary. For a cylinder of radius $R$, for 
example, in cylindrical coordinates $(r,\phi,z)$ with the $z$-axis along the cylinder axis, Eq.\ref{eq:cbc} 
reduces to 
\begin{equation}
\label{eq:bcc}
\left. r \frac{\partial(u_\phi/r)}{\partial r} \right|_{r=R}  = \frac{\Delta V}{\xi}
\end{equation}
where a term $(\partial u_r/\partial\phi)_{r=R}$ is dropped because the cylinder is impenetrable to the fluid. 
If the derivative is expanded, we have
\begin{equation}
        \left.\frac{\partial u_\phi}{\partial r}\right|_{r=R} = {\Delta V \over \Xi} \qquad
{1\over\Xi} \equiv {1\over \xi} + {V\over\Delta V} \cdot {1\over R}
\label{eq:ebc}
\end{equation}
where $V=u_\phi(r=R)$, and in the last equality defined an effective slip length $\Xi$.
If the cylinder is stationary, $\Delta V = V$ and we recover the result of Ref.~\cite{Einzel}. 
However, a more convenient version of the boundary condition follows if we introduce the angular velocity
$\omega(r)=u_\phi(r)/r$ and the angular velocity slip $\Delta \omega= \Delta V/R$:
\begin{equation}
\label{eq:wbc}
	\xi\, \left. \frac{\partial \omega}{\partial r} \right|_{r=R} = \Delta \omega
\end{equation}
Furthermore, these boundary conditions are also valid for spheres. If a sphere of radius $R$ 
rotates about its $z$-axis, we can choose a spherical coordinate system $(r,\theta,\phi)$ oriented 
along this axis and evaluate the strain tensor, and obtain an equation of the same form as 
Eq.~\ref{eq:bcc}, but with a different definition of radius $r$. The boundary condition in the form of 
Eq.~\ref{eq:ebc} follows immediately.  In this case, the fluid's angular velocity with 
respect to the sphere axis is $\omega(r)=u_\phi(r)/(r\sin\theta)$, and if we define the angular velocity 
slip as $\Delta\omega = \Delta V/(R\sin\theta)$ the result is Eq.~\ref{eq:wbc}.

To sum up, we regard the tensor equation Eq.~\ref{eq:gbc} as the general form of the slip boundary 
condition, which is made explicit for a plane surface in Eq.~\ref{eq:lbc} and for a cylindrical or 
spherical surface in Eq.~\ref{eq:wbc}.  The subsequent calculations test whether the single slip length
parameter $\xi$ depends on fluid and solid properties alone.

We employ standard molecular dynamics (MD) techniques \cite{md1, md2, md3} and 
generic interactions of Lennard-Jones form for all interactions between atoms 
\begin{equation}\label{eq:lj}
 V_{ij}(r)=4\epsilon \left[ \left(\frac{\sigma}{r}\right)^{12}
                        - c_{ij}\,\left(\frac{\sigma}{r}\right)^{6}\right]
\end{equation}
The parameter $c_{ij}$ is used to adjust the interaction strength between atomic species 
$i$ and $j$. Here, we have 2 species of atoms in the simulations, fluid (f) and solid (w). 
The solid atoms will constitute the walls in channel flows and the cylindrical and spherical 
particles in other cases.  The interaction coefficient between atoms of the same species is 
always set to unity, $c_{ff}=c_{ww}=1$, while the fluid-solid interaction strength varies to 
adjust the wettability. 
The interaction depth $\epsilon$, the (common) atomic mass $m$ and approximate atomic diameter 
$\sigma$ set the energy, mass and length units, respectively. 
Further details on the MD simulations are given in the Supplementary 
Material \cite{suppl}. 

\begin{figure}[ht]
  	\includegraphics[width=0.45\textwidth]{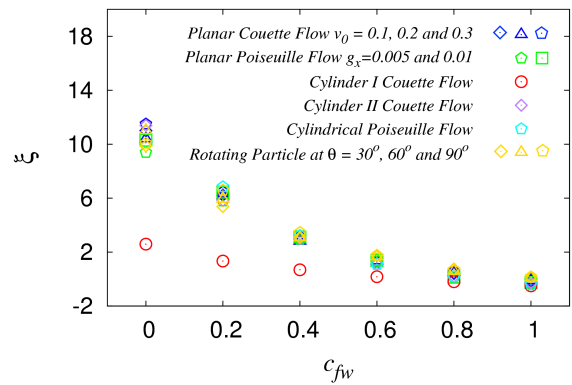}
	\caption{(Color online) Slip length {\em vs}. solid-liquid interaction strength.}
	\label{fig:master}
\end{figure}

The principal result of this paper is the master curve Fig.~\ref{fig:master} for slip length 
{\em vs}. solid-liquid interaction strength, which encapsulates the outcome of all of our simulations 
for flow past planar and curved surfaces.  We now describe the various calculations individually.

\begin{figure}[ht]
        \includegraphics[width=0.38\textwidth]{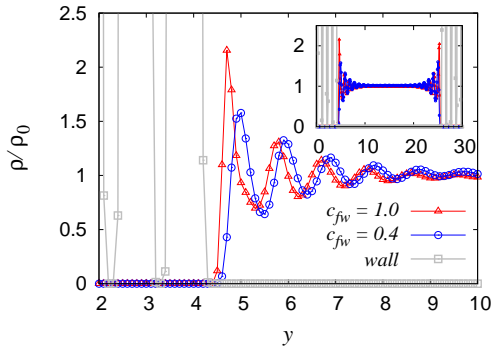}
        \caption{(Color online) Density distribution during planar Couette flow, for two choices
        of interaction strength. }
\label{fig:planar}
\end{figure}

\noindent {\em Plane walls:} We first determine the basic slip length for our model fluid and solid 
systems using
MD simulations of Couette and Poiseuille in a channel between two flat walls, since
we require benchmark values before proceeding to curved surfaces.  
The simulations involve a monatomic liquid between flat parallel walls, the latter consisting of 
layers of solid atoms tethered to cubic lattice sites.  
The position of the ``wall'' must be defined precisely to measure the
slip length, but there is an inherent ambiguity as seen in Fig.~\ref{fig:planar}:
there is a gap between solid and fluid atoms, originating from the repulsive hard core of the potential. 
The nominal or geometrical position of the walls might be taken as 
the $y$-coordinates of the innermost solid layers (as used in \cite{Barrat1}), but we prefer to use 
the midpoint of the gap between the density peak of two phases. The wall position then depends on the
interactions, as does the slip length itself.

To extract the slip length, the measured velocity profiles are fit to the appropriate general 
solution of the Navier-Stokes equation, linear and quadratic functions of $y$ for Couette 
and Poiseuille, respectively, and Eq.~\ref{eq:lbc} is applied.  The result is plotted in
Fig.~\ref{fig:master}, and the numerical values and statistical errors are given in the Supplement
\cite{suppl}.  The slip length is expected to be constant only 
in the Newtonian fluid regime (see, {\em e.g}., \cite{Thompson}) so we 
have also recorded the strain rate for each simulation and 
we note that if limit ourselves to simulations with $\dot{\gamma}\lesssim 0.03$ the slip length is
indeed constant, and it is only these results which 
appear in the figure. The slip length is negligibly small (less than an atomic diameter) for 
strong solid-fluid interaction strength $c_{fw}\ge 0.8$, increases as the interaction weakens, 
and is large but finite as $c_{fw}\to 0$. The special case of {\em no} solid-fluid attraction has
some subtleties: an unconfined liquid drop or film tends to float off or bounce along such a 
surface, and might be said to have infinite slip length, as suggested in \cite{Barrat2}, but a 
dense fluid confined in a channel still experiences some wall friction in the sense that atoms 
near the surface are slowed because their motion is obstructed by the atomic corrugations of 
the surface. The result is a shallow parabolic profile and a finite slip length (and relatively large
statistical fluctuations) rather than a plug flow with infinite $\xi$.

\begin{figure}[ht]
	\includegraphics[width=0.45\textwidth]{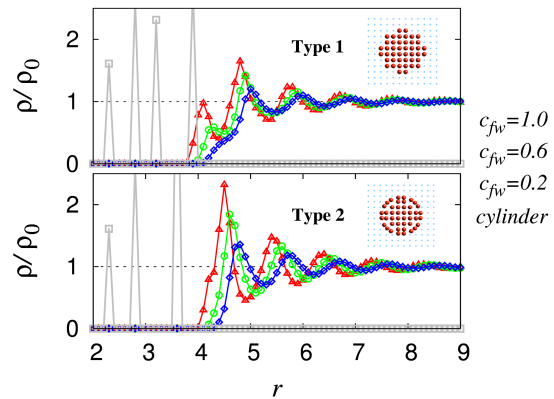}
	\caption{(Color online)  Density profile for the fluid around a rotating 
	type 1 (top) and type 2 (bottom) cylinder, for different values of $c_{fw}$.
 	}
	\label{fg:scd} 
\end{figure}

\noindent {\em Cylinders:} The simplest method to construct a cylinder in MD is to select the
atoms in a cylindrical selection of simple cubic lattice, a ``type 1'' cylinder, as indicated in the 
inset to the left frame of 
Fig.~\ref{fg:scd}. However, the surface is evidently much rougher than a plane surface at the
same atomic density, and its slip properties are likely to differ since experiments indicate a
sensitivity to roughness.  To minimize the effect of surface irregularity, we adjust the atomic 
positions so that each surface atom is at the same radial distance from the cylinder axis, giving 
the smoother ``type 2'' cylinder depicted in the inset to the right frame of the figure, which 
more closely resembles the planar surface treated previously.  
	
The fluid density profile is strongly influenced by the surface structure variation, as seen in  
Fig.~\ref{fg:scd}.  The sharp density peak adjacent to the surface in the planar case is present in
the smoother type 2 cylinder, but the surface interstices in type 1 trap some fluid atoms
and broaden the interfacial region, and make it difficult to assign a single radius to the 
cylinder.  Furthermore, the trapped atoms 
always have a strong attraction to the rest of the fluid, independent of $c_{fw}$, and tend to 
drag it along with the cylinder as it rotates.  Thus, we expect less slip in the type 1 case.  
The general solution of the Navier-Stokes equation for this geometry, assuming cylindrical symmetry
and a velocity which decays at large distances, is $\omega(r) = u_{\phi}(r) /r = k/r^2$, 
where $k$ is a constant.  The measured angular velocity fits this function quite
well, and the boundary condition Eq.~\ref{eq:wbc} determines the slip length.  The results are 
tabulated in \cite{suppl} and included in Fig.~\ref{fig:master}, along with the previous 
slip length as determined from the channel flows. The agreement is excellent for the (smoothed) 
type 2 cylinder, and as expected the slip lengths for the rougher type 1 cylinder are systematically 
lower.  We have verified that the strain rates for these
simulations are sufficiently low as to be in the Newtonian regime.

The finite size of the simulation is a possible source of concern, because periodic boundary
conditions force the fluid velocity to vanish at the edges of the simulation box rather than
decaying to zero as $1/r$.  To test the sensitivity of the results to size we carried out two variant
simulations involving either the same (type 2) cylinder in a smaller box 
or a larger cylinder in the original box:  the results \cite{suppl} are in
agreement with the previous values.
	
\begin{figure}[ht]
        \includegraphics[width=0.45\textwidth]{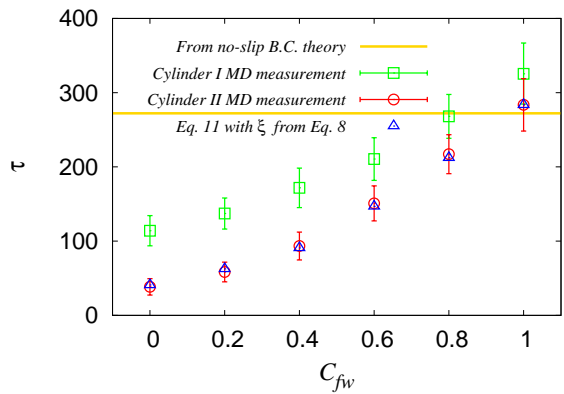}
        \caption{(Color online) Torque {\em vs}. $c_{fw}$ for a cylinder
        rotating with angular velocity $\omega=0.0571$. Simulation results for both cylinder types
        compared to the Navier-Stokes prediction with the slip length determined from Couette
        flow.  The horizontal gold line is the ideal hydrodynamic (no-slip) result.}
        \label{fig:tor}
\end{figure}

We can confirm these results via an independent measurement by determining the 
torque on the rotating cylinder in two ways: first summing the individual torques 
exerted on the cylinder atoms, 
and second by evaluating it from the solution of the Navier-Stokes equation for
a rotating cylinder with a slip boundary condition. Equating the two results determines 
the slip length.  The direct torque measurements are plotted in Fig.~\ref{fig:tor}.
Using the Navier-Stokes solution above and the boundary condition 
Eq.~\ref{eq:wbc} gives $k=u_0R^2/(R+2\xi)$.  The resulting torque is
\begin{equation}
	{\cal{T}}= {4\pi\mu u_0 R^2 L\over R+2 \xi}
\end{equation}
This expression, using the previously determined values for $\xi$ for each $c_{fw}$,
and $\mu=2.2$ as determined from the channel flow simulations for this fluid,
is plotted in Fig.~\ref{fig:tor}, and agrees well with the direct torque measurement for 
the smoothed type 2 sphere.  The slip length values themselves are incorporated in
Fig.~\ref{fig:master} and tabulated in \cite{suppl}.  The torque values measured directly for the 
rough cylinder give larger values (by about 30\%) as one might expect. 

For a variant form of flow along a curved boundary, we consider fluid in the interior of a hollow 
cylinder driven along the axis by a pressure gradient.
The cylinder is made of a section of a cubic lattice by selecting all atoms between an inner and
radius In contrast to the previous case of a thin solid cylinder, 
here the inner radius is large enough that the roughness induced by curvature is insignificant.
In this situation the tensor boundary condition 
Eq.~\ref{eq:gbc} reduces to the simple form $\xi(\partial u_z/\partial r)_{r=R} = \Delta V_z$,
because there is no curvature in the flow direction.
Although there is a superficial resemblance to flow along a flat boundary, the curvature of the wall 
can alter the structure of the nearby fluid layer and there is no guarantee that the slip is the same.   
Simulations of this flow for a range of forcing values in the low strain rate  
regime ($\dot{\gamma} < 0.03$) produce a parabolic velocity profile.
The appropriate general solution of the Navier-Stokes equation (which is regular at $r=0$) is
$u_z=k_1\,r^2+k_2$, and  by fitting the data to this function and applying the boundary condition
we obtain the ``cylindrical Poiseuille flow'' points in Fig.~\ref{fig:master}, again tabulated in 
\cite{suppl}.  The slip lengths are again the same, within statistical uncertainty.

\noindent {\em Spheres} 
Lastly, we turn to the slip characteristics of flow around a spherical particle, a very common 
situation in numerous applications at all length scales. We consider the simplest configuration, a
sphere with a fixed center rotating about a diameter.  As in the cylinder case, the relevant boundary
condition is Eq.~\ref{eq:wbc}  and the same issue of surface roughness arises.  We focus on the
type 2 smooth sphere case alone, where the atoms are first selected from a spherical region of a cubic
lattice, and then those near the boundary are displaced outwards to form a smoothed shell.
The fluid density profile around the sphere resembles that of the smoothed type 2
cylinder discussed above. 
When the sphere rotates at fixed angular velocity $\omega_0$, the 
velocity at the surface varies with polar angle as $\omega_0\sin\theta$, so each angle requires a separate
analysis.  The angular velocity variation with $r$ also resembles the cylinder case, and are fit to
the appropriate Navier-Stokes solution for rotating spheres, $\omega(r)=c_2/r^3$.  

The resulting slip lengths are plotted in Fig.~\ref{fig:master} and tabulated in \cite{suppl}, and once more
the results are consistent with the earlier determinations.  The statistical errors are larger in this
case, up to 8\%, because only a disk-shaped region of the sphere surface is available at a given angle
and the sample is smaller. 

\noindent {\em Conclusion:}
We have used molecular dynamics simulations of flow past stationary surfaces and around rotating solids
to study the variation of slip length with surface curvature.  Provided the slip length is defined in a
consistent tensorial manner, the resulting numerical values depend only on the physical properties of
the solid and fluid involved and do not vary with the flow configuration.  These calculations support
(but of course do not prove) the belief that
the slip length $\xi$ is an intrinsic material property suitable for a fluid mechanical boundary
condition.

The methods of this paper can be applied directly to particles such as ellipsoids, whose shape is 
bounded by coordinate axes in some curvilinear system, rotating about a symmetry axis. For other
rotation axes, or for particles of more complicated shape, the general boundary condition Eq.~\ref{eq:gbc}
is relevant, but an MD analysis along the present lines would be difficult because of sampling issues
-- long runs would be needed to accumulate accurate data on a small surface region.  Likewise,
although this paper is restricted to the low strain rate regime where the slip length is constant, the
methods could easily be extended to that case, but there is no reason to expect the conclusions to
change. Surface roughness on large length scales is not an issue here because the slip boundary
condition is applied locally, but small-scale roughness is problematic.  We have seen that atomic
roughness of the type 1 cylinder changes the slip length significantly, for example.  In this case
approaches involving effective surfaces \cite{Vinogradova} may provide a useful way to characterize slip.

\begin{acknowledgments}
We thank Sepideh Razavi for discussions that prompted this study.
This work is supported in part by the National Science Foundation.
\end{acknowledgments}

\end{document}